\documentstyle[preprint,aps]{revtex}
\newcommand{\cN}{{\cal N}}
\newcommand{\cM}{{\cal M}}
\newcommand{\lmax}{{l_{\rm max}}}

\begin{document}
\draft
\gdef\journal#1, #2, #3, 1#4#5#6{{#1~}{#2 (1#4#5#6) #3}}
\gdef\ibid#1, #2, 1#3#4#5{#1 (1#3#4#5) #2}
\title{Towards a quantum-mechanical model for multispecies
exclusion statistics}
\author{Stefan Mashkevich}
\address{Institute for Theoretical Physics, 252143 Kiev, Ukraine%
\footnote{Present address. Email: mash@phys.ntnu.no} \\
{\rm and} \\
Centre for Advanced Study of the Norwegian Academy of Science
and Letters, \\
Drammensveien 78, N-0271 Oslo, Norway}
\date{\today}
\maketitle
\begin{abstract}
It is shown how to construct many-particle quantum-mechanical
spectra of particles obeying multispecies exclusion statistics,
both in one and in two dimensions.
These spectra are derived from the generalized exclusion
principle and yield the same thermodynamic quantities
as deduced from Haldane's multiplicity formula. \\[1cm]
\end{abstract}

\section{Introduction}
Exclusion statistics was introduced by Haldane \cite{H91}
by postulating a ``generalized exclusion principle'',
according to which adding a particle of species $b$
to a system reduces by $g_{ab}$ the number
of single-particle states available for
each subsequent particle of species $a$.
Thus, the single-particle Hilbert space dimension is
\begin{equation}
D_a=D^{(0)}_a-\sum_b g_{ab}(N_b-\delta_{ab})\;,
\label{D}
\end{equation}
where $D^{(0)}_a$ is that in the absence of particles and
$N_a$ is the number of particles of species $a$;
the many-particle dimension is taken to be
\begin{equation}
W=\prod_a \frac{(D_a+N_a-1)!}{N_a!(D_a-1)!} \;.
\label{W}
\end{equation}
The quantities $g_{ab}$ are called mutual statistics
parameters and constitute the statistics matrix
$G = \{g_{ab} : a,b=1,\ldots,s\}$, where $s$ is the
total number of species.

Starting from these formulas, it is possible
to calculate the thermodynamic quantities:
the distribution function (at least implicitly)
and the equation of state (at least in the form of
a virial expansion);
see Refs.~\cite{W94,IAMP96} for the single-species case
and Refs.~\cite{SS} for the multispecies case.
In particular, when the exponent $\sigma$
in the single-particle dispersion law
$\varepsilon(p)=\varepsilon_0 + \Lambda p^\sigma$
is equal to the dimensionality of space, all the $G$
dependence of the equation of state
is contained in the second order of the virial expansion
\cite{IAMP96,SS,MS94,DVOMS95}.

The question to be addressed here is the following.
What is the many-particle spectrum of noninteracting
particles obeying exclusion statistics in a potential
with a given single-particle spectrum? Understanding
this is an important step towards solving a more general
problem of finding multispecies integrable models that
realize exclusion statistics. Thus far, there are two known
kinds of such models: (i) the single-species one-dimensional
Calogero model \cite{C69} (and the Sutherland model \cite{Suth});
(ii) the single-species \cite{dVO94,dVOMPL95} and multispecies
\cite{SSS,SS} lowest Landau level (LLL) anyon models.
The spectrum of the Calogero model
can be viewed either as that of interacting bosons/fermions
or of free particles obeying exclusion statistics \cite{K93,I94},
the $1/x^2$ interaction being traded
for a ``renormalization'' (fractionization) of quantum numbers,
which can be equivalently regarded
as implied by the generalized exclusion principle.
(In some way, this is reminiscent of the equivalence of
Aharonov--Bohm interaction and anyon statistics \cite{W82}.)
The same structure of the spectrum occurs for the
LLL anyon model, which can be also generalized
to many species.

In the present Letter, we will point out a simple and
natural way of constructing multispecies many-particle
spectra (in a harmonic potential)
for a symmetric statistics matrix $\{g_{ab}\}$,
that are to be viewed as a realization of
exclusion statistics in the sense that they have
the generalized exclusion principle built in and lead
to the thermodynamics implied by Eqs.~(\ref{D}) and (\ref{W}).
After reviewing the one-dimensional case,
we make the construction in two dimensions,
where the underlying idea is to
introduce the fractional exclusion principle
``along one degree of freedom''.

\section{One dimension: Calogero model}
We start with recalling the known
single-species one-dimensional case.
The single-particle energy spec\-trum
in a harmonic potential with frequency $\omega$ is
\begin{equation}
E_n = \left(n + \frac{1}{2}\right)\omega\; ,
\label{en}
\end{equation}
where $n = 0,1,2,\ldots\;$. An $N$-boson state is characterized by
a sequence of quantum numbers $\{ n_j : j=1,\ldots,N \}$ such that
$n_{j+1} \ge n_j$, and its energy is
$E_{ \{n_j\} }(0) = \sum_{j=1}^N E_{n_j}$.
The $N$-boson partition function is
\begin{equation}
Z_N(0)  =  \sum_{\{n_j\}}
\exp [-\beta E_{ \{n_j\} }(0)]
 = \frac{\exp[N^2x/2]} {\prod_{j=1}^N [\exp(jx)-1]} \; ,
\label{zbos}
\end{equation}
where $x=\beta\omega$.
The spectrum of the Calogero model,
\begin{equation}
E_{ \{n_j\} }(g) = \left[ \sum_{j=1}^N \left(n_j + \frac12\right)
+ \frac{N(N-1)}{2}g \right]\omega \; ,
\label{ecal}
\end{equation}
where $g(g-1)$ is the coupling constant of the inverse square interaction,
can be obtained in terms of free particles obeying exclusion
statistics with parameter $g$, in the following way:
Regard $n$ in Eq.~(\ref{en}) as a continuous variable
and postulate that every particle $k$ ``pushes'' every particle $j$
residing higher than $k$, by $g$ units up, in terms of $n$.
This is understood as $g$ states being excluded
for particle $j$.
Mathematically, this means that the
quantum numbers are ``renormalized'':
$n_j \to \tilde n_j$, with
\begin{equation}
\tilde n_j = n_j + \sum_{k=1}^N \theta(\tilde n_j - \tilde n_k) g \; ,
\label{nj}
\end{equation}
where $\theta$ is the step function.
With $n_{j+1} \ge n_j$, the solution to this equation is
$\tilde n_j = n_j + (j-1)g$ \cite{K93};
that is, the lowest particle remains in its place, the second lowest one
gets shifted by $g$ units up, the third one by $2g$ up, etc.
On the other hand, there is no interaction now, so
the many-body energy is a sum
\begin{equation}
E_{ \{n_j\} }(g) = \sum_{j=1}^N E_{\tilde n_j} \; .
\end{equation}
By summing (\ref{nj}) over $j$, one sees that
renormalizing the $n_j$'s increases the energy by $g\omega$
per each pair of particles.
Hence, Eq.~(\ref{ecal}) follows immediately.
In particular, for $g=1$ one obtains the fermionic spectrum.
The partition function is
$ Z_N(g) = \exp \left[ -\frac12N(N-1)gx \right] Z_N(0) $,
and the thermodynamics is the same \cite{IAMP96,SS,MS94,DVOMS95}
as obtained starting from Eqs.~(\ref{D})--(\ref{W}).

\section{A multispecies generalization}
A state of $s$ species of bosons is characterized
by a set of quantum numbers
$\{n_{aj}: a=1,\ldots,s, \; j=1,\ldots,N_a\}$
with $n_{a,j+1}\ge n_{aj}$, its energy is
$E_{ \{n_{aj}\} }(0) =
\sum_{a=1}^s \sum_{j=1}^{N_a} E_{n_{aj}}$,
and the partition function is
\begin{equation}
Z_{N_1\ldots N_s}(0) = \prod_{a=1}^s Z_{N_a}(0) \; .
\label{zmultibos}
\end{equation}

We will assume the statistics matrix
to be symmetric: $g_{ab} = g_{ba}$---an
important restriction, which will be discussed later.
The appropriate generalization of (\ref{nj}) is (cf.~\cite{IV96})
\begin{equation}
\tilde n_{aj} = n_{aj} + \sum_{b=1}^s \sum_{k=1}^{N_b}
\theta(\tilde n_{aj} - \tilde n_{bk}) g_{ab} \; .
\label{nmulti}
\end{equation}
This means that a particle of species $b$ ``pushes'' one of
species $a$ by $g_{ab}$ units up whenever the latter resides
higher than the former in the {\it final} state. Obviously,
in contrast to the single-species case, the order of $\tilde n$'s
may not coincide with that of $n$'s, so that Eq.~(\ref{nmulti})
has to be solved in a self-consistent manner.
However, assuming again the free-particle expression for the energy
\begin{equation}
E_{ \{n_{aj}\} }(G) =
\sum_{a=1}^s \sum_{j=1}^{N_a} E_{\tilde n_{aj}}\;,
\label{emulti}
\end{equation}
the form of Eq.~(\ref{nmulti}) implies that, just like in the
single-species case, each pair of particles, one of species $a$
and the other one of species $b$, contributes $g_{ab}\omega$
to the energy. Hence,
\begin{equation}
E_{ \{n_{aj}\} }(G) =
\Bigg[ \sum_{a=1}^s \sum_{j=1}^{N_a} \left( n_{aj} + \frac12 \right)
+ \sum_{a=1}^s \frac{N_a(N_a-1)}{2}g_{aa}
+ \sum_{\scriptstyle a,b=1 \atop \scriptstyle a<b}^s
N_a N_b g_{ab} \Bigg] \omega \; .
\label{emultical}
\end{equation}

This is essentially the same as the LLL spectrum of
multispecies anyons (with charges of the same sign) \cite{SSS,SS}.
Moreover, a semiclassical treatment of that problem \cite{SS}
leads to Eq.~(\ref{nmulti}). The problem itself is two-dimensional, but
the LLL restriction renders it effectively one-dimensional
by reducing the phase space. In the genuine one-dimensional case,
however, it is not known what Hamiltonian would be
a multispecies generalization of the Calogero one
and possess the spectrum (\ref{emultical}).
At least two possible such generalizations, one
\cite{FO95} involving and the other one \cite{S96}
not involving a three-particle interaction, were
considered in the literature,
but neither of those serves our purpose.

\section{Thermodynamics}
Be that as it may, Eq.~(\ref{emultical}) by itself is enough
to derive the thermodynamic functions.
{}From the grand partition function
$\Xi = \sum_{N_1\ldots N_s} z_1^{N_1} \cdots z_s^{N_s} Z_{N_1\ldots N_s}$
[where $z_a = \exp(\beta \mu_a)$], one can deduce the cluster expansion
$\ln \Xi = \sum_{k_1\ldots k_s} b_{k_1\ldots k_s} z_1^{k_1} \cdots z_s^{k_s}$
and the dimensionless analog of the virial expansion
$\ln \Xi = \sum_{k_1\ldots k_s} A_{k_1\ldots k_s}
N_1^{k_1} \cdots N_s^{k_s}/Z_1^{k_1+\cdots+k_s-1}$,
using the relation $N_a = z_a (\partial \ln \Xi/\partial z_a)$;
the coefficients $A_{k_1\ldots k_s}$ are just numbers.
If there is a well-defined volume,
the usual virial expansion (the equation of state)
$\beta P = \sum_{k_1\ldots k_s} a_{k_1\ldots k_s}
\rho_1^{k_1} \cdots \rho_s^{k_s}$ follows, by
$\ln \Xi = \beta PV, \;\; N_a = \rho_a V$.

The partition function that one obtains from Eq.~(\ref{emultical}) is
\begin{equation}
Z_{N_1\ldots N_s}(G) =
\exp \Bigg[ -\Bigg(\sum_{a=1}^s \frac{N_a(N_a-1)}{2}g_{aa}
+ \sum_{\scriptstyle a,b=1 \atop \scriptstyle a<b}^s
N_a N_b g_{ab}\Bigg) x \Bigg] Z_{N_1\ldots N_s}(0) \; .
\label{Zmulti}
\end{equation}
\sloppy
This yields the cluster coefficients in the thermodynamic limit,
\begin{eqnarray}
&& b_{k_1\ldots k_s}(G) = F_{k_1\ldots k_s}(G) x^{-1}\;,
\label{clus1d} \\
&& F_{k_1\ldots k_s}(G) = (-1)^{s-1}
\frac{\prod_{j=1}^s \prod_{l=1}^{k_j-1}
\left[ 1 - (\sum_{n=1}^s k_n g_{jn})/l \right]}
{\prod_{n=1}^s k_n^2}
\sum_{{p_1,q_1,\ldots, \atop p_{s-1},q_{s-1}=1} \atop
p_n < q_n}^s \prod_{j=1}^{s-1} k_{p_j} k_{q_j} g_{p_jq_j} \; ;
\label{f}
\end{eqnarray}
in the last sum, all the pairs $(p_j, q_j)$ must be distinct,
and each number from $1$ to $s$ must figure in the set
$\{ p_1,\ldots,p_{s-1},q_1,\ldots,q_{s-1} \}$ at least once.
All the numbers $k_1, \ldots , k_s$
are assumed to be different from zero
(otherwise, an obvious invariance with respect to the
renumbering of the species and the identity
$F_{k_1\ldots k_r0\ldots0} =
F_{k_1\ldots k_r}$ should be used).
For example:
\begin{eqnarray*}
F_{400}(G) & = & \frac{1}{16}
(1 - 4g_{11})(1 - \frac42g_{11})(1 - \frac43g_{11})\;; \\
F_{321}(G) & = & \frac{1}{6}
\left[ 1 - (3g_{11} + 2g_{12} + g_{13}) \right]
\left[ 1 - \frac12(3g_{11} + 2g_{12} + g_{13})\right]
\left[ 1 - (3g_{12} + 2g_{22} + g_{23})\right] \\
&& {} \times
\left( 3g_{12}g_{13} + 2g_{12}g_{23} + g_{13}g_{23} \right) \;.
\end{eqnarray*}

To compare, the general formula for particles obeying exclusion
statistics in a $D$-dimensional harmonic potential,
implied by Eqs.~(\ref{D}) and (\ref{W}), is \cite{SS}
\begin{equation}
b_{k_1\ldots k_s}(G) =
(k_1+\cdots+k_s)^{-D} f_{k_1\ldots k_s}(G)\, Z_1 \;,
\label{clusgen}
\end{equation}
and for a symmetric $G$ it turns out that
$f_{k_1\ldots k_s}(G) = (k_1+\cdots+k_s) F_{k_1\ldots k_s}(G)$.
Since in the thermodynamic limit $Z_1=1/x$, the system at hand
does exhibit exclusion statistics.
The dimensionless virial coefficients are
\begin{eqnarray}
A_{2} & = & (2g_{11}-1)/4 \; ,
\qquad A_{11} = g_{12} \; ; \nonumber \\
A_{k} & = & {\cal B}_{k-1}/k! \quad \mbox{for}\; k>2 \; ,
\label{am}
\end{eqnarray}
and all the others vanish. Thus, they only depend on
the statistics at the second order \cite{IAMP96,SS,MS94,DVOMS95}.

The system considered is, in fact, thermodynamically equivalent
to a system of particles with linear dispersion
$\varepsilon(p) = \Lambda p$ in a box
(because the single-particle spectra are equidistant
in both cases, and the renormalization will alter them
in the same way)
of a volume related to the harmonic frequency
in such a way that the single-particle
partition functions $Z_1$ in both cases match:
$1/x = V/\pi\Lambda\beta$,
or $V = \pi \Lambda /\omega$.
The virial coefficients of the latter system are
$a_{k_1\ldots k_s} = A_{k_1\ldots k_s}
(\pi\Lambda\beta)^{k_1+\cdots+k_s-1}$.

\section{Two dimensions}
The single-particle spectrum in a two-dimensional
harmonic potential is
\begin{equation}
E_{mn} = (m + n + 1)\omega \;,
\end{equation}
where $m,n = 0,1,2,\ldots\;$.
The quantum numbers $m$ and $n$ correspond to the two
degrees of freedom (say, $x$ and $y$ motion).
The idea is to implement the fractional exclusion
principle ``along one degree of freedom'',
i.e., with respect to one quantum number.
Thus, split up the spectrum into ``sectors'',
labeled by $m$, each sector being the
spectrum of a one-dimensional
oscillator with a constant $(m+1/2)\omega$ added to the levels,
and postulate exclusion statistics in such a way:
Particles within one sector behave like they do in the above
multispecies one-dimensional model, while particles in different
sectors do not influence one another.

Formally, it looks this way.
A state of $s$ species of bosons can be characterized by a set of pairs
$\{ (m_{aj}, n_{aj}): a=1,\ldots,s, \; j=1,\ldots,N_a \}$
such that either $m_{a,j+1} > m_{aj}$ or ($m_{a,j+1} = m_{aj}$
and $n_{a,j+1} \ge n_{aj}$). That is, the next particle of
the same species is put either into a higher sector than the
previous one or into the same sector but not lower.
Define a set of numbers $\{ \cN_{al}: a=1,\ldots,s, \; l=1,\ldots,\lmax \}$,
where $\cN_{al}$ is the number of particles of species $a$ in the
$l$th from below nonempty sector and $\lmax$ is the total number
of nonempty sectors. One has $N_a = \sum_{l=1}^\lmax \cN_{al}$.
Introduce now a quantity $\cM_{al} = \sum_{l'=1}^l \cN_{al'}$;
then for any $a$, the $l$-th nonempty sector
contains those particles $(aj)$ for which
$\cM_{a,l-1}+1 \le j \le \cM_{al}$.
Eq.~(\ref{nmulti}), in accordance with the aforesaid,
is replaced by
\begin{equation}
\tilde n_{aj} = n_{aj} + \sum_{b=1}^s \sum_{k=1}^{N_b}
\theta(\tilde n_{aj} - \tilde n_{bk}) \delta_{m_{aj}m_{bk}} g_{ab} \; ,
\label{nmulti2d}
\end{equation}
whereas $\tilde m_{aj} = m_{aj}$.
The free-particle expression for the energy
\begin{equation}
E_{ \{(m_{aj},n_{aj})\} }(G) = \sum_{a=1}^s \sum_{j=1}^{N_a}
E_{\tilde m_{aj} \tilde n_{aj} }
\end{equation}
now yields
\begin{eqnarray}
&& E_{ \{(m_{aj},n_{aj})\} }(G) = \sum_{l=1}^\lmax
\Bigg[ \sum_{a=1}^s
\cN_{al} \left( m_l+\frac{1}{2} \right)
+ \sum_{a=1}^s \sum_{j=\cM_{a,l-1}+1}^{\cM_{al}}
\left( n_{aj}+\frac12 \right)
\nonumber \\ && \quad
{} + \sum_{a=1}^s \frac{\cN_{al}(\cN_{al}-1)}{2}g_{aa}
{} + \sum_{\scriptstyle a,b=1 \atop \scriptstyle a<b}^s
\cN_{al} \cN_{bl} g_{ab} \Bigg] \omega \; ,
\label{e2dim}
\end{eqnarray}
where $m_l$ is the absolute number of the $l$th nonempty sector, so
that $m_{aj}=m_l$ with $l$ such that $\cM_{a,l-1}+1 \le j \le \cM_{al}$.
The quantity under the outermost sum can be recognized as
Eq.~(\ref{emultical}) applied within the $l$-th sector, plus
a constant displacement determined by $m_l$.
For example, let in the case of Bose statistics there be
particles of species 1 in the states (0,1), (2,4), (3,1);
particles of species 2 in the states (0,0), (2,1), (2,2);
particles of species 3 in the states (1,2), (2,2), (3,2).
The energy of the corresponding state
with exclusion statistics will be
$(3+g_{12})+(4)+(21+2g_{12}+g_{13}+g_{22}+2g_{23})+(11+g_{13})
=39+3g_{12}+2g_{13}+g_{22}+2g_{23}$.

The partition function takes the form
\begin{eqnarray}
&& Z_{N_1\ldots N_s}(G) = \sum_{ \{ \cN_{al} \} }
\sum_{m_1=0}^\infty \sum_{m_2=m_1+1}^\infty \cdots
\sum_{m_\lmax=m_{\lmax-1}+1}^\infty
\prod_{l=1}^\lmax \Bigg\{
\left( \prod_{a=1}^s Z_{\cN_{al}}(0) \right)
\nonumber \\ && \quad
{} \times \exp \Bigg[ - \Bigg( \sum_{a=1}^s \cN_{al}
\left( m_l+\frac12 \right)
+ \sum_{a=1}^s \frac{\cN_{al}(\cN_{al}-1)}{2}g_{aa}
+ \sum_{\scriptstyle a,b=1 \atop \scriptstyle a<b}^s
\cN_{al} \cN_{bl} g_{ab}
\Bigg) x \Bigg] \Bigg\} \; . \label{z2d}
\end{eqnarray}
The outermost sum is over all possible distributions of particles
over sectors, with $\lmax$ determined by such a distribution.

The cluster coefficients that one obtains from this are
\begin{equation}
b_{k_1\ldots k_s}(G) =
(k_1+\cdots+k_s)^{-1} F_{k_1\ldots k_s}(G) x^{-2}
\label{clus2dim}
\end{equation}
with $F_{k_1\ldots k_s}(G)$ as in Eq.~(\ref{f}).
This is but again the general formula (\ref{clusgen})
for $D=2$, so that exclusion statistics is present.
The virial coefficients do not exhibit any specific pattern.

It is possible, then, to obtain the cluster coefficients
of the same system in a box of volume $V$ rather than in
a harmonic potential \cite{SS,SSS,Comt89,Ol92}:
\begin{equation}
b^V_{k_1\ldots k_s}(G) = \frac{Vx^2}{\lambda^2}
(k_1+\cdots+k_s)b_{k_1\ldots k_s}(G)
\label{clusosc}
\end{equation}
(the dispersion law being quadratic, $\varepsilon(p) = p^2/2m$,
and $\lambda^2 = 2\pi \beta/m$).
These coincide, up to a common factor, with the ones from
Eq.~(\ref{clus1d}), and the dimensionless
virial coefficients are again those given by Eq.~(\ref{am}).
(In fact, they will be the same for any system with exclusion
statistics with a constant single-particle density of states.)
The dimensional virial coefficients are
$a_{k_1\ldots k_s}(G) = A_{k_1\ldots k_s}(G)
\lambda^{2(k_1+\cdots+k_s-1)}$.

\section{Discussion and conclusions}
Essentially, what has been shown here is that
(i) for a one-dimensional harmonic potential,
``excluding $g_{ab}$ (single-particle) states''
as required by the generalized exclusion principle
has to be understood as
renormalizing the quantum numbers
so that the sum of those for a pair of particle $a$
and particle $b$ increases by $g_{ab}$;
(ii) in two dimensions, the same procedure has to
be followed within effectively one-dimensional
subsets of the spectrum, corresponding to one degree
of freedom. In fact, we have also verified that the same
thermodynamics is obtained when in the two-dimensional
problem, the sectors are made up of states with the
same angular momentum, which appears to be a better choice
since it preserves rotational invariance.
It is a plausible assumption that any division by
sectors is actually good, and that the same will be
true even for the three-dimensional problem
[where $E_{lmn} = (l+m+n+\frac{3}{2})\omega$ and the sectors
would be labeled by $l$ and $m$].

It is, however, not clear at this point
how to construct the spectrum if the matrix
$g_{ab}$ is not symmetric. In deriving
Eq.~(\ref{emultical}), or (\ref{e2dim}), it was
implied that $\theta(x-y) + \theta(y-x) = 1$
always. However, putting $\theta(0) = 1/2$
and using it for a nonsymmetric matrix does not
lead to the same thermodynamics as obtained
from Eqs.~(\ref{D}) and (\ref{W}).
Perhaps some generalization of
Eq.~(\ref{nmulti}), or (\ref{nmulti2d}),
is necessary in this case.
Apparently, there is a relevant physical example:
Interpreting numerically found few-electron spectra
in terms of fractional quantum Hall effect quasiparticles and
comparing the multiplicities
to Haldane's formula \cite{antisym} seems to hint
that the system of two species of anyons with opposite charges
in the LLL should exhibit exclusion statistics with $g_{12}=-g_{21}$.
To prove (or disprove) this directly by finding the spectrum,
i.e., by generalizing Eq.~(\ref{emultical}),
remains an open problem.

For a symmetric statistics matrix, both the spectrum
and the thermodynamics of the conjectured
integrable models realizing multispecies exclusion
statistics are now known, which provides one with
an essential information helping to search for such models.

\section*{Acknowledgements}
I would like to thank Serguei Isakov for many
useful discussions and Diptiman Sen for
a number of comments. I am grateful to the Centre for
Advanced Study in Oslo, where this work was initiated,
for kind hospitality and financial support.
The analytic formulas for the cluster and virial
coefficients were derived using {\it Mathematica} \cite{Wolf}.

\end{document}